\documentclass{book}    % fundamental class file is 'book'.
\usepackage{piers}  % use the file 'piers.sty'.
\begin{document}

%=== Input Title here ================================
\title{Dissipative breathers in rf SQUID metamaterials}
\maketitle
%===================================================

%=== List of authors (in order) ========
%-- Author(s) for the first affiliation ---
\author      {F. M. Lastname}
\affiliation {University}
\address     {}% optional
\city        {Boston}
\postalcode  {}% optional
\country     {USA}
\phone       {}    % optional
\fax         {}    % optional
\email       {email@email.com}  % optional
\misc        { }  % optional
\nomakeauthor
%------------------------------------

%=== List of authors (in order) ========
%-- Author(s) for the second affiliation ---
\author      {F. M. Lastname}
\affiliation {University}
\address     {}% optional
\city        {Boston}
\postalcode  {}% optional
\country     {USA}
\phone       {}    % optional
\fax         {}    % optional
\email       {email@email.com}  % optional
\misc        { }  % optional
\nomakeauthor
%-------------------------------------
%=== List of other authors (in order) ========
%......

%---Output of Authors----------------------
\begin{authors}

{\bf G. P. Tsironis}$^{1}$, {\bf N. Lazarides}$^{1,2}$, {\bf and M. Eleftheriou}$^{1,3}$\\
\medskip
$^{1}$Department of Physics, University of Crete,
and Institute of Electronic Structure and Laser,
Foundation for Research and Technology-Hellas,
P. O. Box 2208, 71003 Heraklion, Greece \\
$^{2}$Department of Electrical Engineering,
Technological Educational Institute of Crete,
P. O. Box 140, Stavromenos, 71500, Heraklion, Crete, Greece\\
$^{3}$Department of Music Technology and Acoustics,
Technological Educational Institute of Crete,
E. Daskalaki, Perivolia,
74100 Rethymno, Crete, Greece
\end{authors}
%--------------------------

%---Content of Paper Abstract-----------------------
\begin{paper}

\begin{piersabstract}
The existence and stability of dissipative breathers in
rf SQUID (Superconducting Quantum Interference Device)
arrays is investigated numerically. 
In such arrays, the nonlinearity which is intrinsic to each SQUID,
along with the weak magnetic coupling of each SQUID to its nearest neighbors,
result in the formation of discrete breathers.
We analyze several discrete breather excitations in rf SQUID arrays
driven by alternating flux sources in the presence of losses.
The delicate balance between internal power losses and input power,
results in the formation of 
dissipative discrete breather (DDB) structures up to relatively
large coupling parameters.
It is shown that DDBs may locally alter the magnetic response of an rf
SQUID array from paramagnetic to diamagnetic or vice versa.
\end{piersabstract}

%---Content of Paper Text-----------------------
\psection{Introduction}
The discrete breathers (DBs), which are also known as intrinsic localized modes (ILMs),
belong to a class of nonlinear excitations that appear generically
in discrete and spatially extended systems \cite{Flach}.
They are loosely defined as spatially localized, time-periodic and stable
excitations, that can be produced spontaneously in a nonlinear lattice
of weakly coupled elements as a result of fluctuations \cite{Peyrard},
disorded \cite{Rasmussen},
or by purely deterministic mechanisms \cite{Hennig}.
The last two decades, a large number of theoretical and experimental studies
have explored the existence and the properties of DBs in a variety of
nonlinear discrete systems. Nowadays, there are rigorous mathematical proofs
of existence of DBs both for energy conserved and dissipative systems
\cite{Mackay,Aubry}, and several numerical algorithms
for their accurate construction have been proposed \cite{Marin,Marin1}.
Moreover, 
they have been observed experimentally in a variety of systems,
including solid state mixed-valence transition metal complexes \cite{Swanson},
quasi-one dimensional antiferromagnetic chains \cite{Schwarz},
arrays of Josephson junctions \cite{Trias},
micromechanical oscillators \cite{Sato},
optical waveguide systems \cite{Eisenberg},
layered crystal insulator at $300 K$ \cite{Russell},
and proteins \cite{Edler}.

From the perspective of applications to experimental situations
where an excitation is subjected to dissipation and external driving,
dissipative DBs (DDBs) are more relevant than their
energy conserved counterparts.
The dynamics of DDBs is governed by a delicate balance
between the input power and internal power losses.
Recently, DDBs have been demonstrated numerically in discrete
and nonlinear magnetic metamaterial (MM) models
\cite{Lazarides,Eleftheriou}.
The MMs are artificial composites that exhibit electromagnetic (EM) properties
not available in naturally occuring materials.
They are typically made of subwavelength resonant elements like, for
example, the split-ring resonator (SRR).
When driven by an alternating EM field,
the MMs exhibit large magnetic response, either positive or negative,
at frequencies ranging from the microwave up to the Terahertz and
the optical bands \cite{Linden,Shalaev}.
The magnetic response of materials at those frequencies is particularly
important for the implementation of devices such as compact cavities,
tunable mirrors, isolators, and converters.
The nonlinearity offers the possibility 
to achieve dynamic control over  the response of a metamaterial in real time,
and thus tuning its properties by changing the intensity of the external field.
Recently, the construction of nonlinear SRR-based MMs \cite{Shadrivov} 
gives the opportunity to  test experimentally the existence of DDBs
in those materials.

It has been suggested that periodic rf SQUID arrays can operate
as nonlinear MMs in microwaves,
due to the resonant nature of the SQUID itself and the nonlinearity that 
is inherent to it \cite{Lazarides2}.
The combined effects of nonlinearity and discreteness (also inherent
in rf SQUID arrays), may lead in the generation of nonlinear excitations
in the form of DDBs \cite{Lazarides1}.
In the present work we investigate numerically the existence and stability 
of DDBs in rf SQUID arrays. In the next section we shortly describe rf SQUID
array model, which consists a simple realization of a planar MM.
In section 3 we present several types of DDBs that have been constructed
using standard numerical algorithms, and we discuss their
magnetic response. We finish in section 4 with the conclusions.
%%%----------figure1---------
\begin{figure}[!t]
\includegraphics[angle=0, width=.6\linewidth]{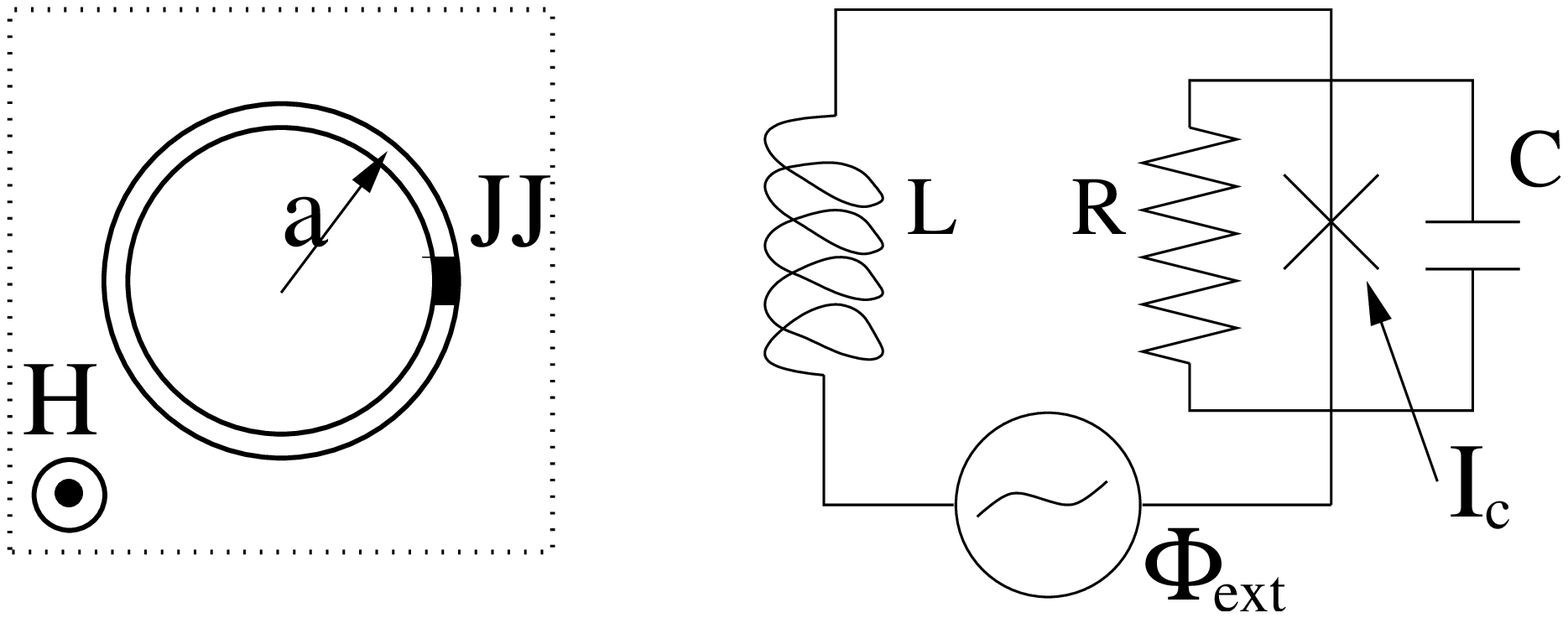}
\hspace{15mm}
\includegraphics[angle=0, width=.25\linewidth]{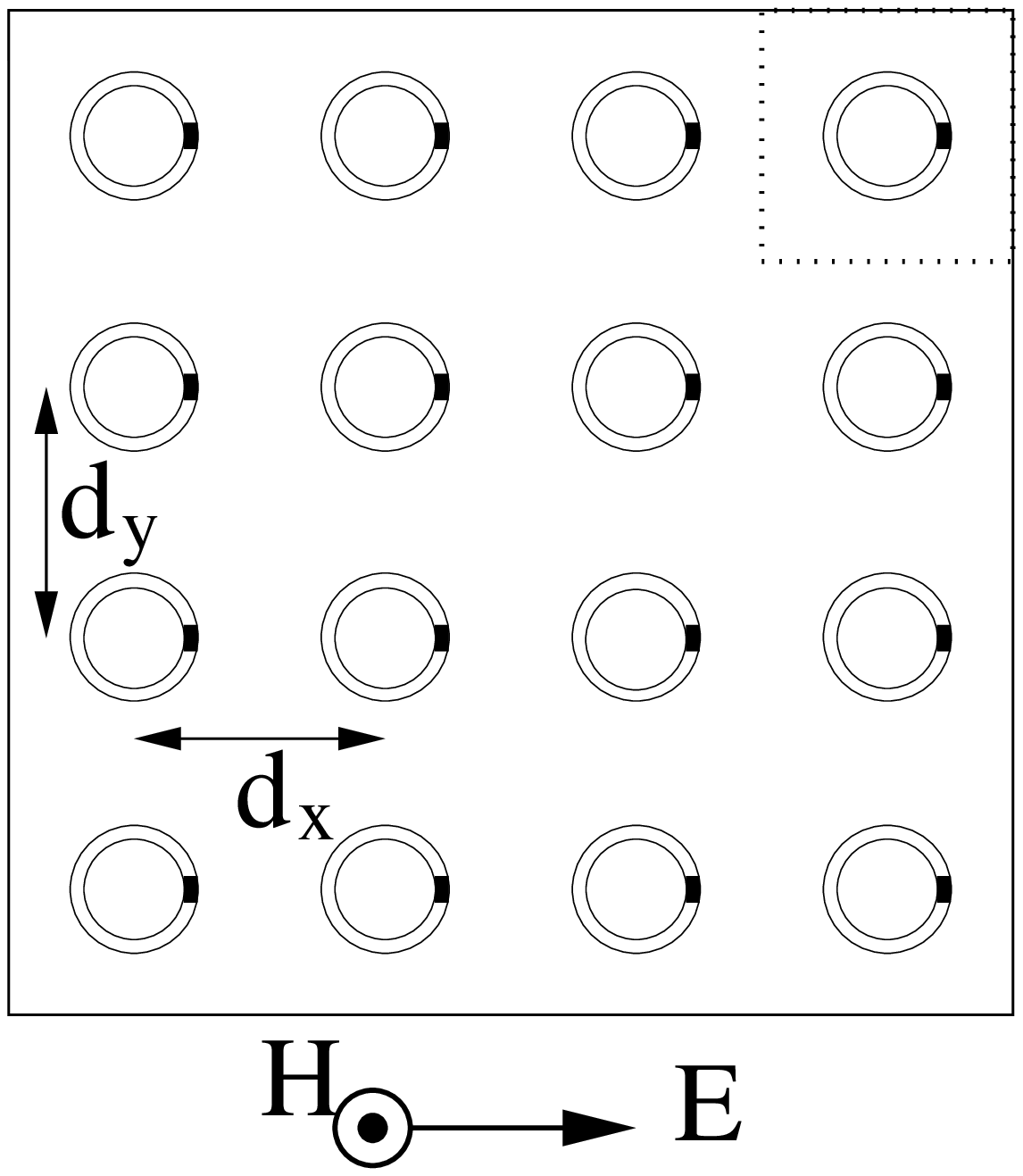}
\caption{
Left panel: Schematic drawing of a ring-shaped rf SQUID.
Middle panel: Equivalent circuit for an rf SQUID in an alternating
magnetic field.
Right panel: Schematic drawing of a two-dimensional orthogonal array of 
identical rf SQUIDs.
}
\end{figure}

\psection{rf SQUID metamaterial model}
An rf SQUID, shown schematically in the left panel of Fig. 1,
consists of a superconducting ring interrupted by a Josephson junction (JJ)
\cite{Likharev}.  When driven by an alternating magnetic field, the induced
supercurrents in the ring are determined by the JJ through the Josephson
relations.  Adopting the resistively and capacitively shunted
junction (RCSJ) model for the JJ \cite{Likharev},
an rf SQUID in an alternating field $H_{ext} \equiv H$ perpendicular
to its plane is equivalent to the lumped circuit model shown in the middle 
panel of Fig. 1.
That circuit consists of an inductance $L$ in series with an ideal Josephson
element $I_c$
(i.e., for which $I=I_c \sin\phi$,
where $I_c$ is the critical current of the JJ and $\phi$ is the Josephson phase)
shunted by a capacitor $C$ and a resistor $R$, driven by an alternating
flux $\Phi_{ext} (H)$.

Consider a planar rf SQUID array consisting of identical units (right panel of Fig. 1),
arranged in an orthogonal lattice with constants $d_x$ and $d_y$ in the
$x$ and $y$ directions, respectively.
That system is placed in a uniform magnetic field
$H=H_{DC} +H_{AC} \, \sin(\omega t)$, where $\omega$ is the frequency and
$t$ is the temporal variable, perpendicular to the SQUID rings.
The field induces a supercurrent $I_{nm}$ in the $nm-$th SQUID through the flux
$\Phi_{ext} = \Phi_{DC} +\Phi_{AC} \, \sin(\omega t)$ threading the SQUID loop
($\Phi_{DC,AC}=\mu_0 S H_{DC,AC} \omega$ is the external flux amplitude, 
with $\mu_0$ being the permeability of the vacuum and $S$ the loop area of the SQUID).
The supercurrent $I_{nm}$ produces a magnetic field which couples that SQUID 
with its first neighbors in the $x$ and $y$ directions, due to magnetic interactions
through their mutual inductances $M_x$ and $M_y$, respectively.
The dynamic equations for the (normalized)
fluxes $f_{nm}$ can be written in the form \cite{Lazarides1}
\begin{eqnarray}
  \label{1}
   \frac{d^2 f_{nm}}{d\tau^2} +\gamma \frac{d f_{nm}}{d\tau} +f_{nm}
    +\beta\, \sin( 2 \pi f_{nm} )
    - \lambda_x ( f_{n-1,m} + f_{n+1,m} )
   - \lambda_y ( f_{n,m-1} + f_{n,m+1} )
   \nonumber \\
     = [1-2(\lambda_x +\lambda_y)] f_{ext} ,
\end{eqnarray}
where the following relations have been used
\begin{equation}
  \label{1a}
    \tau=\omega_0 \, t,~~ \omega_0 =1/\sqrt{L\, C}, ~~
    f_{nm} = \Phi_{nm} / \Phi_0,~~ f_{ext} = \Phi_{ext} / \Phi_0,~~
    \beta= \beta_L / 2\pi \equiv L I_c/\Phi_0 .
\end{equation}
In the earlier equation, $\Phi_0$ is the flux quantum, 
$\beta_L$ is the SQUID parameter, 
$\gamma$ is the dissipation constant,
and $\lambda_{x,y}$ are the
coupling coefficients in the $x$ and $y$ directions,
defined as $\lambda_{x,y} = M_{x,y}/L$, respectively.
The time derivative of $f_{nm}$ corresponds to the voltage $v_{nm}$
across the JJ of the $nm-$th rf SQUID, i.e., $v_{nm} = {d f_{nm}}/{d\tau}$.
The normalized external flux $f_{ext}$ is given by
\begin{eqnarray}
 \label{2}      
   f_{ext} = f_{DC}  + f_{AC} \cos(\Omega \tau ) ,
\end{eqnarray}
where $f_{AC} = \Phi_{AC}/\Phi_0$, $f_{DC} = \Phi_{DC}/\Phi_0$,
and $\Omega =\omega/\omega_0$, 
with $\Phi_{DC}$ being a constant (DC) flux resulting from the 
time-independent component of the magnetic field $H$.

The dispersion for small amplitude flux waves is obtained
by the substitution of
$f= A\, \exp[i (\kappa_x n + \kappa_y m - \Omega \tau)]$,
into the linearized Eqs. (\ref{1}) for $\gamma=0$ and $f_{ext} =0$,
which gives
\begin{eqnarray}
 \label{3}
    \Omega_{\bf \kappa} = \sqrt{1 + \beta_L -2( \lambda_x \, \cos \kappa_x
      +\lambda_y \, \cos \kappa_y ) }  ,
\end{eqnarray}
where ${\bf \kappa}=(\kappa_x, \kappa_y) = (d_x \, k_x, d_y \, k_y)$.
The corresponding one-dimensional (1D) SQUID array is obtained by setting
$\lambda_y=0$, $\lambda_x=\lambda$, $\kappa_x=\kappa$, 
and by dropping the subscript $m$ in Eqs. (\ref{1}).
Typical dispersion curves $\Omega(\kappa)$ for the 1D system are shown 
in Fig. 2a for three different values of the coupling $\lambda$.
The bandwidth $\Delta\Omega \equiv \Omega_{max} - \Omega_{min}$
decreases with decreasing $\lambda$ which leads, 
for $\lambda << 1$ \cite{Kirtley},
to a nearly flat band with $\Delta\Omega \simeq 2 \lambda \sqrt{1+\beta_L}$
(and relative bandwidth $\Delta\Omega /\Omega \simeq 2 \lambda$).
Importantly, the group velocity $v_g$, which defines the
direction of power flow, is in a direction opposite
to the phase velocity $v_{ph}$, as it is observed in Fig. 2b.
%%%----------figure2---------
\begin{center}
\begin{figure}[!t]
\includegraphics[angle=-0, width=.7\linewidth]{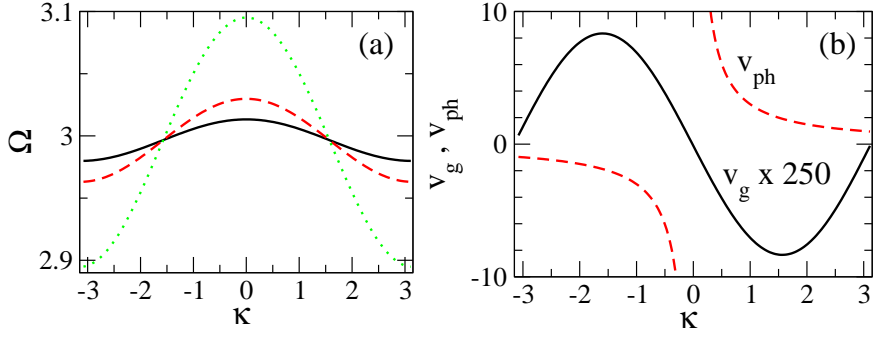}
\caption{
(a) Frequency band $\Omega$ as a function of $\kappa$ for a 1D rf SQUID array,
for $\beta=1.27$, and $\lambda=-0.05$ (narrowest band, black-solid curve),
$\lambda=-0.1$ (red-dashed curve),
$\lambda=-0.3$ (widest band, green-dotted curve).
(b) Group velocity $v_g$ (black-solid curve)
and phase velocity $v_{ph}$ (red-dotted curve), for a 1D rf SQUID array
with $\beta=1.27$ and $\lambda=-0.1$.
}
\end{figure}
\end{center}
%%%----------figure3---------
\begin{figure}[!h]
\includegraphics[angle=-0, width=.5\linewidth]{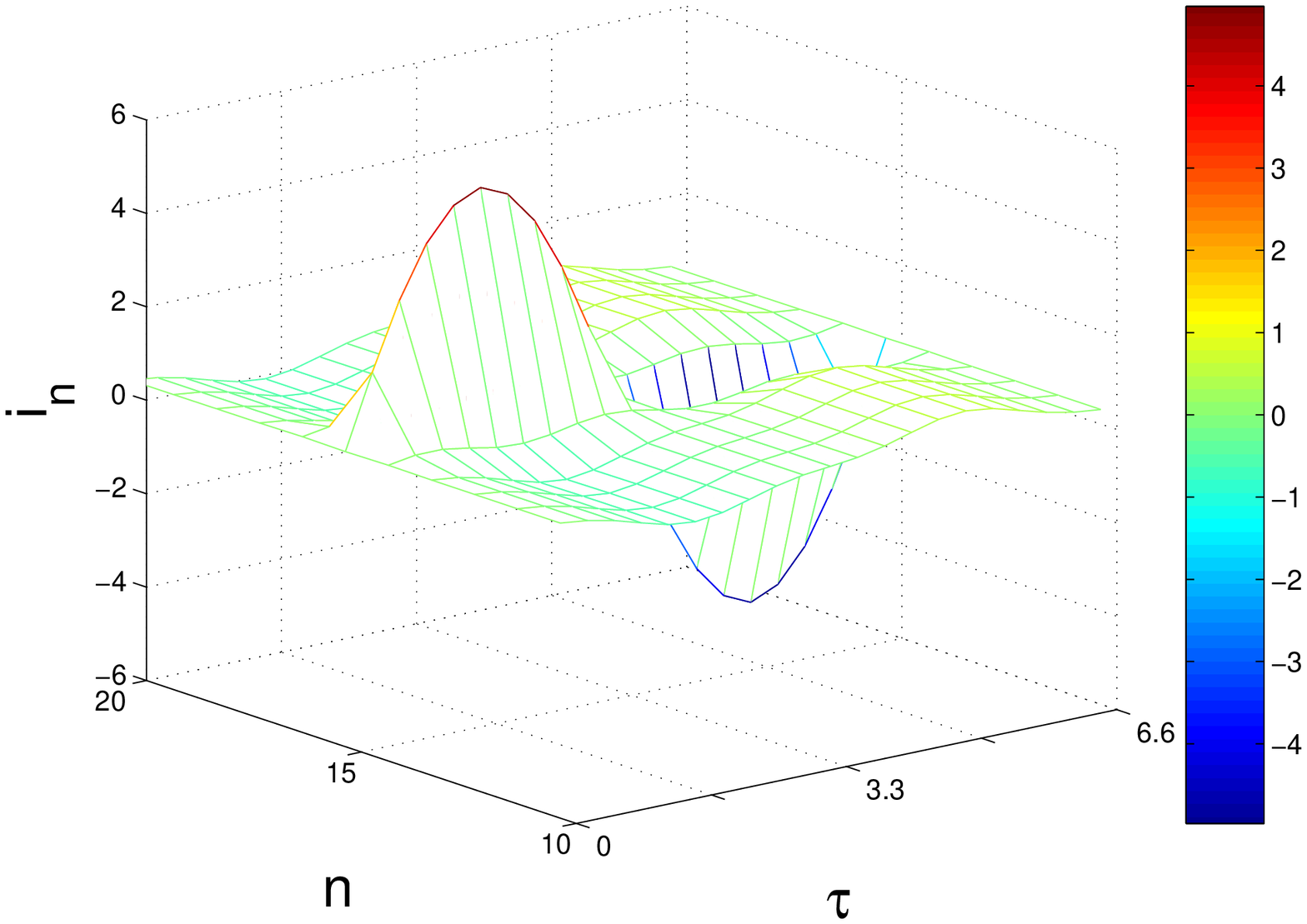}
\includegraphics[angle=-0, width=.5\linewidth]{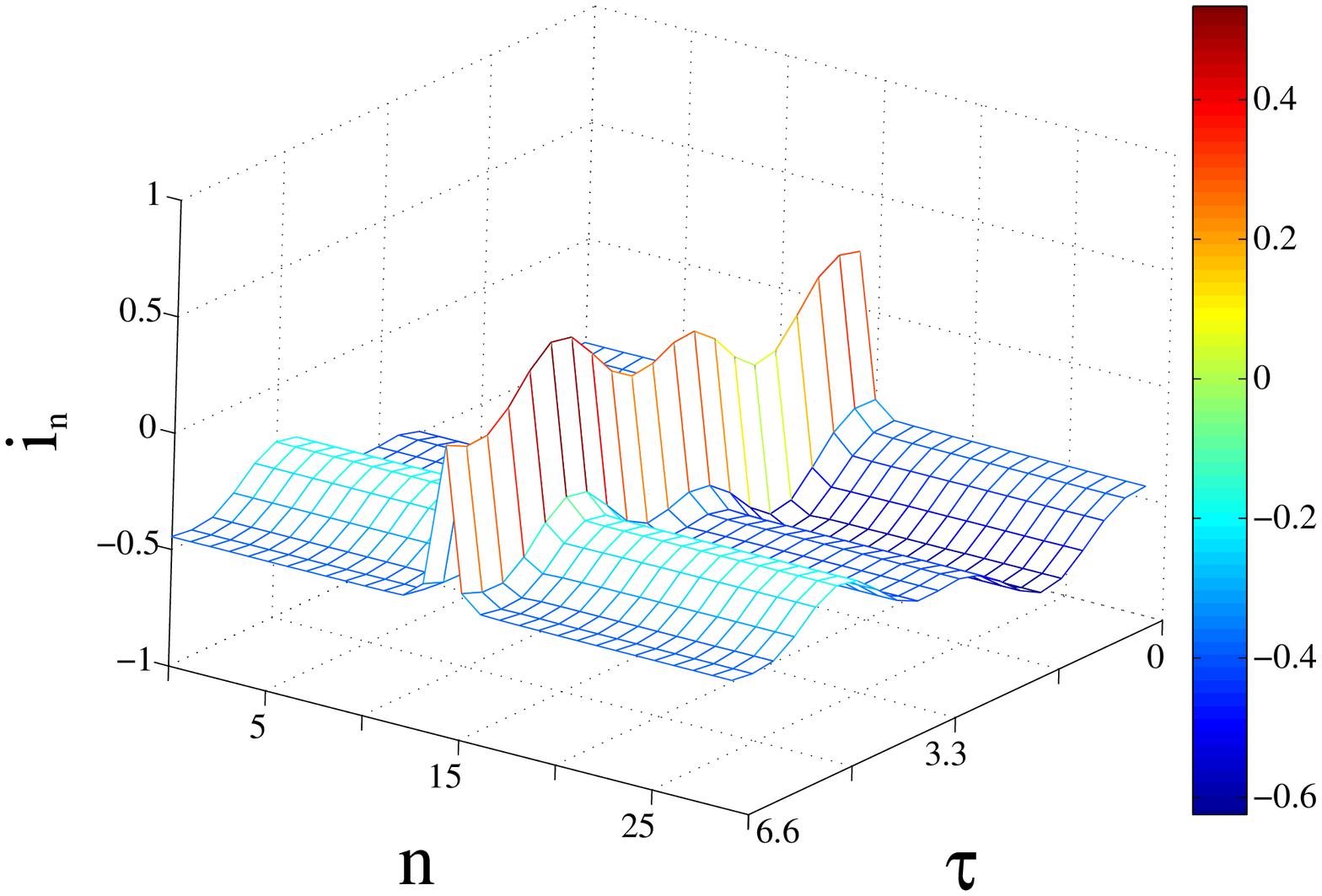}
\caption{
Time evolution of dissipative breathers during one period,
for $\lambda=-0.1$, $T_b=6.6$, $\gamma=0.001$, $\beta=1.27$, and
(right panel) $f_{DC}=0.5$, $f_{AC}=0.2$ - low-amplitude breather;
(left panel) $f_{DC}=0$, $f_{AC}=0.6$ - high-amplitude breather.
Only part of the array ($N=30$) is shown for clarity.
}
\end{figure}

\psection{Dissipative breathers and magnetic response}
For the generation of DDBs in rf SQUID arrays,
we use the algorithm developed by Marin {\em et. al.} \cite{Marin}.
With that algorithm, we can construct low- and high-amplitude
DDBs up to some maximum value of the coupling, $\lambda_{max}$,
which generally depends on the external flux amplitudes $f_{AC}$ and 
$f_{DC}$ \cite{Lazarides1}.
Both the central site and the background of those DDBs are oscillating
with frequency $\Omega_b =2\pi/T_b =\Omega$, i.e., the same as that 
of the external flux, $\Omega$. 
Typical single-site bright DDBs of both low- and high-amplitude 
are shown in Fig. 3 (right and left panels, respectively), where 
the  spatio-temporal evolution of the induced currents $i_n$ ($n=1,2,3,...,N$)
are shown during one DDB period $T_b$.
We should note the non-sinusoidal time-dependence of the oscillations
in both panels of Fig. 3.
The linear stability of DDBs is addressed
through the eigenvalues of the Floquet matrix (Floquet multipliers).
A DDB is linearly stable when all its Floquet multipliers
$m_i, ~i=1,...,2N$, lie on a circle of radius $R_e = \exp(-\gamma T_b/2)$
in the complex plane. The DDBs shown in Fig. 3 are indeed linearly stable.
Moreover, those DDBs were let to evolve for large time intervals
(i.e., more than $10^5~T_b$) without any observable change in their shapes.
With the same algorithm, we can also construct 2D dissipative breathers.
A snapshot of such a DDB taken at maximum amplitude
of the central site is shown in the left panel of Fig. 4. 

The normalized flux through the $nm-$th SQUID can be casted in the form
\begin{equation}
 \label{4}
    \beta \, i_{nm} = f_{nm}^{loc} - f_{ext}^{eff} ,
\end{equation}
where
\begin{eqnarray}
  \label{5}
   f_{nm}^{loc} =f_{nm}
    -\lambda_x ( f_{n-1,m} + f_{n+1,m} )
    -\lambda_y ( f_{n,m-1} + f_{n,m+1} ) , \qquad 
     f_{ext}^{eff} =[1 -2 (\lambda_x + \lambda_y)] f_{ext} .
\end{eqnarray}
%%%----------figure4---------
\begin{figure}[!t]
%%\vspace{1cm}
\includegraphics[angle=-0, width=.4\linewidth]{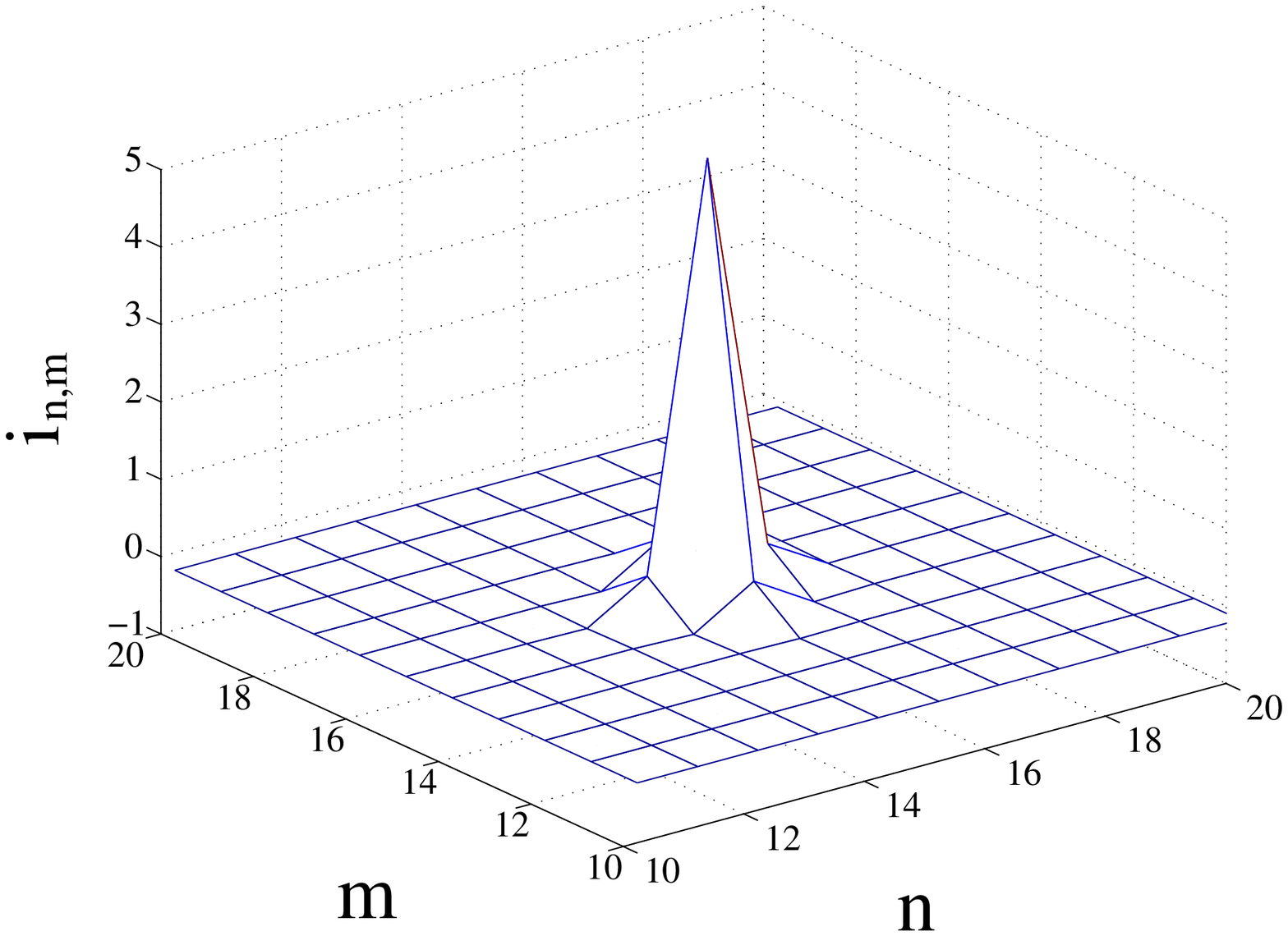}
\hspace{2cm}
\includegraphics[angle=-0, width=.4\linewidth]{Tsironis_PIERS09_Fig4R.eps}
\caption{
Left panel: A snapshot of a two-dimensional dissipative discrete breather (DDB)
for $\lambda_x=\lambda_y=-0.1$ and the other parameters as in the left
panel of Fig. 3.
Right panel:
Temporal evolution of $\beta\, i_n$ (red-solid curve),
$f_n^{loc}$ (green-dashed curve), and $f_{ext}$ (black-dotted curve)
during one period $T_b$, for
(a) the central site of the DDB shown in the left panel ($n=m=n_b=N/2$); 
(b) the site with $n=m=7$ of the DDB shown in the left panel. 
}
\end{figure}
After division by the area of the unit cell $d^2$ of the 2D array,
the terms $f_{ext}^{eff}$, $f_{nm}^{loc}$, and
$\beta \, i_{nm}$ in Eq. (\ref{4}) can be interpreted as the effective
external field,
the local magnetic induction at the $nm-$th cell, and the magnetic response
at the $nm-$th cell, respectively.
The temporal evolution of $\beta \, i_{nm}$, $f_{nm}^{loc}$,
and the external field $f_{ext}$, are shown in the right panel of Fig. 4, 
for two different sites of the 2D DDB shown in the left panel of Fig. 4:
the central DDB site at $n=m=n_b=N/2$, and the site located at $n=m=7$
(Figs. (a) and (b) of the right panel of Fig. 4, respectively).
We observe that in the cell corresponding to the central DDB site
the magnetic response is in phase with the applied field providing a 
strong paramagnetic response, while in the cell corresponding to the site located 
in the background the magnetic response is in anti-phase with the applied field
providing moderate diamagnetic response.
Thus, the local magnetic induction is sharply peaked at the central DDB site,
as can be inferred by comparing the green-dashed curves in (a) and (b) in the 
right panel of Fig. 4.

\psection{Conclusion}
In conclusion, we have shown using standard numerical methods that 
periodic rf SQUID arrays in an alternating external flux support 
low- and high-amplitude linearly stable DDBs.
Those DDBs are not destroyed by increasing the dimensionality  from one to two.
Thus, we have constructed several linearly stable DDB excitations both 
for 1D and 2D  rf SQUID arrays, which may alter locally the magnetic 
response of the arrays. 
Planar SQUID arrays similar to those described here have been actually constructed
and studied with respect to the ground state ordering of their magnetic moments
\cite{Kirtley}. Thus, the above theoretical predictions are experimentally testable.

\end{paper}
%--------------------------

\end{document}